\documentclass[aps,prl,twocolumn,superscriptaddress,floatfix]{revtex4}
\usepackage{amsmath,amssymb,graphicx,graphics,color}
\usepackage{dcolumn}% Align table columns on decimal point
\usepackage{bm}% bold math
\usepackage{psfrag}
\usepackage[T1]{fontenc}
\usepackage{calligra}

\begin{document}

\title{Fidelity Susceptibility Study of Quantum Long-Range Antiferromagnetic Ising Chain}
\author{Gaoyong Sun}
\affiliation{College of Science, Nanjing University of Aeronautics and Astronautics, 211106, China}

\begin{abstract}
We study the fidelity susceptibility of quantum antiferromagnetic Ising chain with a long-range power law interaction $1/r^{\alpha}$ 
using the large-scale density matrix renormalization group method. We find that the critical adiabatic dimension $\mu=2$ 
and the critical exponent of the correlation length $\nu=1$ for arbitrary $\alpha>0$, indicating all quantum phase transitions are second-order Ising transitions. 
In addition, we numerically determine the complete phase diagram for $0 < \alpha \le 3$ from the data collapse of the fidelity susceptibility and show that the critical point 
$h_c$ changes monotonously with respect to $\alpha$. This work will shed light on the nature of phase transitions in the quantum long-range antiferromagnetic Ising chain 
from a quantum information perspective.
\end{abstract}

\maketitle

%%%%%%%%%%%%%%%%%%%%%%%%%%%%%%%
% Introduction
{\it Introduction.-} Quantum many-body models with nearest neighbor interactions, such as the well-known quantum Ising model, the Heisenberg model and the Hubbard model,
are fundamentally important in understanding quantum phase transitions and collective behaviors of strongly correlated quantum many-body systems \cite{Sachdev1999}. 
Besides the nearest neighbor interactions, there are also many types of long-range interactions, e.g., the Coulomb interaction $1/r$ \cite{Saffman2010}, 
the dipole-dipole interaction $1/r^3$ \cite{Lahaye2009}, and the van der Waals-London interaction $1/r^6$ \cite{Saffman2010}.
Quantum many-body models with power-law long range interaction are particular interesting
because of the frustrations and the novel critical behaviors due to the tunable long range interactions, which has motivated 
plenty of experimental and theoretical studies during the last 40 years \cite{ Fisher1972,Dutta2001,Laflorencie2005,Dalmonte2010,Sandvik2010,Schachenmayer2013,
Gong2014, Vodola2014, Angelini2014, Gori2015, Cevolani2015, Gong2016, Maghrebi2016, Santos2016, Kovacs2016, Humeniuk2016, Lepori2016R2, 
Regemortel2016, Buyskikh2016, Bermudez2017, Valiente2017,  Maghrebi2017, Behan2017, Gong2017R2, Jurcevic2017, Halimeh2016,
Lepori2016, Hess2017, Defenu2016, Homrighausen2017,Ho2017,Deng2017,Droennner2017, Barros2017, Pagano2017}. 

Recently, such tunable power-law interaction $1/r^{\alpha}$ with an exponent $0\le \alpha \le 3$ is realized in trapped ions \cite{Britton2012,Islam2013,Richerme2014,Jurcevic2014}, 
which provides a perfect platform for investigating the novel physics of quantum many-body systems in the presence of a power-law interaction $1/r^{\alpha}$
and has stimulated many follow-up works studying the corresponding many-body physics.
However, as far as we know, most of the previous theoretical and numerical studies investigated one-dimensional (1D) quantum many-body models 
with a long-range interaction because of the complexity of two or three dimensional many-body systems although the experimental systems range from one to three dimensions.
In the following, we will focus our discussions on one dimensional quantum long-range antiferromagnetic Ising (LRAI) model \cite{Koffel2012,Hauke2013,Vodola2016,Fey2016,Jaschke2017}
that we will investigate throughout the paper.

\begin{figure}%[ht]
\includegraphics[width=8.6cm]{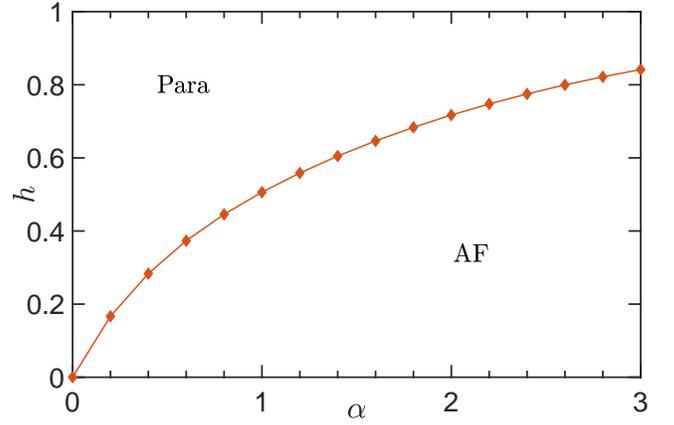}
\caption{(Color online) Phase diagram of the long-range antiferromagnetic Ising chain with respect to $\alpha$ and transverse field $h$ from 
the data collapse of the fidelity susceptibility $\chi_{F}$ for L = 96, 144, 192, 240 sites. AF denotes the antiferromagnetic phase, Para denotes the paramagnetic phase; 
symbols denote the DMRG results of the critical values $h_c$ except the exact result $\alpha=0, h_c=0$ point.}
\label{figPhasediagram}
\end{figure}

The short range transverse field quantum Ising chain ($\alpha=0$) is an exactly solvable model with the critical adiabatic dimension $\mu=2$ 
and the critical exponent of the correlation length $\nu=1$. However when the long-range interactions are included the model becomes 
much more complicated due to the geometric frustration so that it is very difficult to handle it with theoretical treatments or numerical simulations.
Thanks to the development of the density matrix renormalization group (DMRG) technique \cite{White1992,Schollwock2005} , 
the ground-state wave functions and the long-range Hamiltonian can be expressed in terms of matrix product states \cite{Verstraete2004,Schollwock2011}
and the matrix product operators \cite{Crosswhite2008,Pirvu2010} respectively, which makes possible the numerical simulations of the systems. 
Recently, the LRAI chain has been investigated using matrix product states \cite{Koffel2012,Vodola2016,Jaschke2017}, 
where the phase diagram was determined by the maximum of entanglement entropy. An analytical tool called linked-cluster expansions (LCEs) were used 
for the LRAI chain to obtain the phase diagram and the critical exponents \cite{Fey2016}.

Studies showed that the system undergoes a second order phase transition from an antiferromagnetic order to a paramagnetic order for all $\alpha$. 
However, the nature of such second order phase transitions is still under debate especially for smaller $\alpha <1$ \cite{Koffel2012,Vodola2016,Fey2016,Jaschke2017}.
In this letter, we investigate the nature of the phase transitions for the first time by studying the fidelity susceptibility in the long-range antiferromagnetic Ising chain 
with a power-law interaction $1/r^{\alpha}$ using the large-scale DMRG method. 
By very carefully computing the ground-state wave functions, we obtain the critical adiabatic dimension $\mu$, the critical exponent of the correlation length $\nu$
and the ground-state phase diagram by the finite-size scaling and the data collapse of fidelity susceptibility. Surprisingly we find that all the second-order 
quantum phase transitions for $0 < \alpha \le 3$ are Ising transitions with the Ising universality class $\mu=2, \nu=1$ in contrast to previous results \cite{Koffel2012,Vodola2016}
and the corresponding critical point $h_c$ changes monotonously with respect to $\alpha$ that is perfectly consistent with the results in Ref. \cite{Koffel2012,Vodola2016}.

\begin{figure}%[ht]
\includegraphics[width=8.6cm]{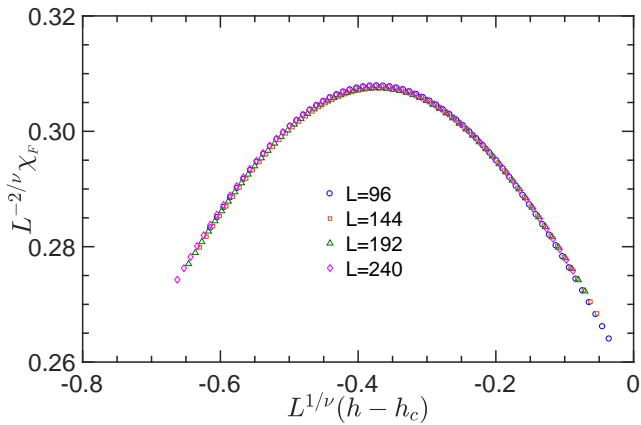}
\caption{(Color online) Data collapse of fidelity susceptibility $\chi_{F}$ for the long-range antiferromagnetic Ising chain; 
symbols denote the DMRG results for $\alpha=0.6$ and $L=96,144,192,240$ sites, where $\nu=1.008$ and $h_c=0.3733$ are used for the data collapse plots.}
\label{collapsedata}
\end{figure}

%%%%%%%%%%%%%%%%%%%%%%%%%%%%%%%
% Model
{\it Model.-} The quantum long-range antiferromagnetic Ising (LRAI) chain is given by \cite{Britton2012,Islam2013,Richerme2014,Jurcevic2014,
Koffel2012,Hauke2013,Vodola2016,Fey2016,Jaschke2017}

\begin{align}
H ={}& J\sum_{i<j} \frac{\sigma^{z}_{i}\sigma^{z}_{j}}{|i-j|^{\alpha}} - h \sum_{i}\sigma^{x}_{i}
\label{model}
\end{align}
where $\sigma^z_{i}$ and $\sigma^x_{i}$ are Pauli matrices at the $i$th site, $J$ denotes the strength of the interaction and $h$ represents the transverse field.
$\alpha>0$ is a parameter to tune the range of the power-law interactions $r^{-\alpha}$ with $r=|i-j|$. For example, if $\alpha=0$, the model is an infinite-range Ising chain.
Other important well-known interactions are Coulomb-like interaction ($\alpha=1$), dipole-dipole interactions ($\alpha=3$) and van der Waals interaction ($\alpha=6$).
Since there is no exact solution for the LRAI chain of arbitrary $\alpha$, we employ the DMRG method \cite{White1992,Schollwock2005} 
based on matrix product states \cite{Verstraete2004,Schollwock2011}, which is one of the most powerful unbiased numerical methods 
for one dimensional strongly correlated  many-body systems. The long-range power-law interactions $r^{-\alpha}$ is approximated in terms of
exponentially decaying interactions \cite{Crosswhite2008,Pirvu2010}
\begin{align}
r^{-\alpha}=\sum_{j}a_{j}b^{r-1}_{j}
\end{align}
for a given $\alpha$. The coefficients $a_j$ and $b_j$ are found by minimizing the distance using the nonlinear least-squares method
\begin{align}
\sum_{j=1}^{N} \sum_{r=1}^{r_m}(a_{j}b^{r-1}_{j}-r^{-\alpha})^2
\end{align}
The maximal absolute error of each site is controlled within $10^{-8}$ up to $r_m=300$ sites.
The long-range Hamiltonian $r^{-\alpha}$ is then captured by a matrix product operator (MPO) with a virtual dimension $3N+2$.
We choose the strength of the interaction $J=1$ and use open boundary conditions throughout the paper.

\begin{figure}%[ht]
\includegraphics[width=8.7cm]{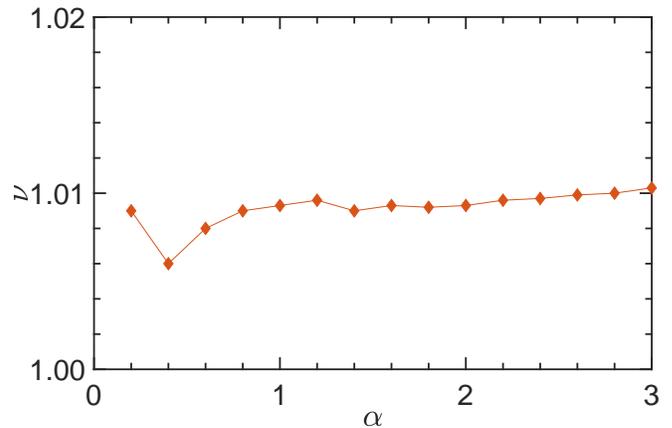}
\caption{(Color online) Critical exponent of the correlation length $\nu$ with respect to $\alpha$ for the long-range antiferromagnetic Ising chain;
symbols denote the DMRG results of $\nu$ that were obtained by the data collapse from the fidelity susceptibility $\chi_{F}$ 
for $L=96,144,192,240$ sites.}
\label{nudata}
\end{figure}

%%%%%%%%%%%%%%%%%%%%%%%%%%%%%%%
% Fidelity Susceptibility 
{\it Fidelity susceptibility.-} Given a Hamiltonian $H(h)=H_0+hH_1$ with the driving parameter $h$, the quantum ground-state fidelity
\cite{Zanardi2006,Venuti2007,You2007} $F(h,h+\delta h)$ is the overlap amplitude of two ground states $|\psi(h) \rangle$ and $|\psi(h+\delta h) \rangle$
\begin{align}
F(h,h+\delta h)=| \langle \psi(h) | \psi(h+\delta h) \rangle |
\end{align}
It is believed that the fidelity $F(h,h+\delta h)$ can characterize the quantum phase transition \cite{Venuti2007,Kwok2008,You2007,Schwandt2009,Albuquerque2010,Gu2010,Greschner2013,Sun2016} 
due to the qualitative differences of the ground state $|\psi(h) \rangle$ at the phase transition point $h=h_c$.
Alternatively, a more convenient quantity - fidelity susceptibility $\chi_{F}(h)$ that appears as the leading quadratic term in the expansion of ground-state fidelity $F(h,h+\delta h)$ is defined as 
\begin{align}
\chi_{F}(h)=\lim_{\delta h \rightarrow 0} \frac{-2\ln F(h,h+\delta h)}{(\delta h)^2}
\label{eqFS}
\end{align}
A large number of works have been done to investigate the relation between fidelity susceptibility and quantum phase transitions including the second order phase transitions \cite{Gu2010}
and topological Berezinsky-Kosterlitz-Thouless (BKT) transitions \cite{Chen2008,Yang2007,Fjærestad2018,Langari2012,Carrasquilla2013,Lacki2014,Wang2015,Sun2015}. 
Studies \cite{Gu2010} show that for a second-order quantum phase transition, 
the finite-size scaling behavior of fidelity susceptibility $\chi_F(h)$ in the vicinity of the quantum critical point $h_c$ is
\begin{align}
\chi_{F}(h \rightarrow h_c) \propto  L^{\mu}
\label{eqFSscaling}
\end{align}
where $L$ is the size of system and $\mu$ is the critical adiabatic dimension.
On the other hand, it is shown in Ref.(\cite{Albuquerque2010}) that the fidelity susceptibility per site scales as
\begin{align}
L^{-d} \chi_{F}(h) = L^{(2/v)-d}f_{\chi_{F}}(L^{1/\nu}|h-h_c|)
\label{eqFScollapse}
\end{align}
from the scaling analysis for a second-order quantum phase transition. Where $f_{\chi_{F}}$ is a unknown scaling function,
$\nu$ is the critical exponent of the correlation length and $d$ is the dimension of the system. 
Hence it provides a very simple approach to determine the universality class of a quantum phase transition by finding 
the critical adiabatic dimension $\mu$ and the critical exponent of the correlation length $\nu$.
For example, both exact and numerical results show that $\mu=2$ and $\nu=1$ for quantum Ising model with short-range interaction \cite{Zanardi2006,Gu2010,Damski2013,Damski2014}.
The fidelity susceptibility $\chi_{F}(h)$ can be either an intensive quantity for a gapped system or a super extensive quantity for a gapless system \cite{Gu2010}. 
Therefore, in practice the fidelity susceptibility per site 
\begin{align}
\chi_{L}(h)=\chi_{F}(h)/L^d
\label{eqFSsite}
\end{align}
is usually used for extracting the critical exponent $\mu$. 

\begin{figure}%[ht]
\includegraphics[width=8.6cm]{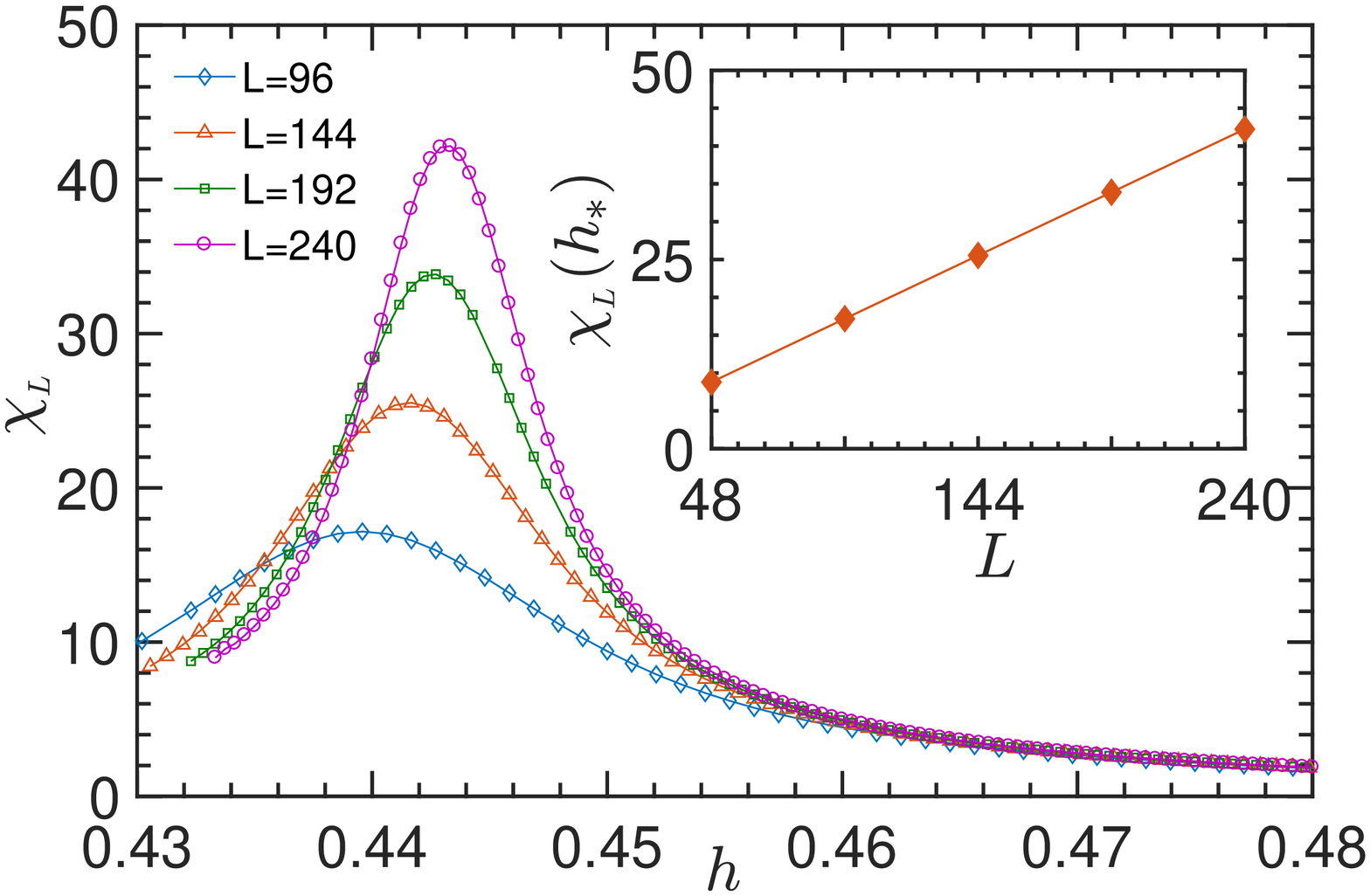}
\caption{(Color online) Fidelity susceptibility per site $\chi_L$ of the long-range antiferromagnetic Ising chain for $\alpha=0.8$ and $L=96,144,192,240$ sites
as a function of the transverse field $h$; symbols denote DMRG results. Insert shows the finite-size scaling 
of the amplitude of fidelity susceptibility per site $\chi_{L}(h_\ast)$ at the peak position $h_{\ast}$.}
\label{FSdata}
\end{figure}

%%%%%%%%%%%%%%%%%%%%%%%%%%%%%%%
% Numerical Results
{\it Phase diagram.-} There are two limiting cases: $\alpha = 0$ and $\alpha \rightarrow \infty$ for the LRAI chain. For $\alpha=0$, it is the Ising chain with an infinite-range interaction.
And at $h=0$ point, the phase is classical with infinite degenerate ground states. The antiferromagnetic N{\'e}el phase is one of the ground states. 
Infinitezimal quantum fluctuations coming from field $h$ lift the degeneracy immediately, resulting in a paramagnetic phase for $h>0$.
At $h=0, \alpha>0$ the classical ground state becomes the antiferromagnetic N{\'e}el phase, with only 2-fold degeneracy protected by Z2 symmetry.
For $\alpha \rightarrow \infty$, the model is a standard short-range Ising chain, the model has two gapped phases 
(antiferromagnetic phase and paramagnetic phase) at the critical point $h_c=1$. 

\begin{figure}%[ht]
\includegraphics[width=8.7cm]{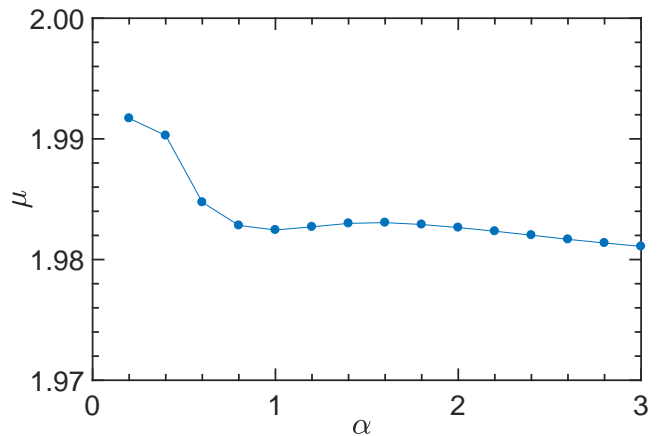}
\caption{(Color online) Critical adiabatic dimension $\mu$ with respect to $\alpha$ for the long-range antiferromagnetic Ising chain;
the symbols denote the DMRG results of $\mu$ that were obtained by extrapolating from the fidelity susceptibility $\chi_{F}(h_\ast)$ 
at the peak position $h_{\ast}$ of $L=96,144,192,240$ sites.}
\label{Mudata}
\end{figure}

To determine the full phase diagram of LRAI chain, we perform the DMRG calculations for $L=48, 96,144,192,240$ sites and $M=500$ states. 
We compute the fidelity susceptibility as defined in Eq.(\ref{eqFS}) with the minimal step $\delta h=10^{-4}$. 
The phase diagram is plotted in Fig.\ref{figPhasediagram} for $0 < \alpha \le 3$ as a function of $\alpha$ and $h$.
The critical value $h_c$ (and also the critical exponent of the correlation length $\nu$) are determined by the data collapse of the fidelity susceptibility,
which is achieved by plotting the scaling function $L^{-2/\nu}\chi_{F}(h)$ in Eq.(\ref{eqFScollapse}) with respect to $L^{1/v}(h-h_c)$. 
The accuracy of the critical value $h_c$ and the critical exponent of the correlation length $\nu$ is the order of $10^{-4}$.

As an example, we plot the scaling function $L^{-2/\nu}\chi_{F}(h)$ as a function of $L^{1/v}(h-h_c)$ for $\alpha=0.6$ and $L=96,144,192,240$ sites shown in Fig.\ref{collapsedata}, 
where the critical value $h_c=0.3733$ and the critical exponent $\nu=1.008$ are found to be the best fitting values for the data collapse.
We found that the critical value $h_c$ changes monotonously with respect to $\alpha$ and saturate to $h_c=1$, which is consistent with the previous numerical 
results \cite{Koffel2012,Vodola2016,Fey2016,Jaschke2017} and the exact solution of short-range Ising model in the limiting case $\alpha \rightarrow \infty$. 
Interestingly, the critical exponent of the correlation length $\nu$ as shown in Fig.\ref{nudata} are found to be $\nu=1$ belonging to 
the Ising universality class for arbitrary $\alpha>0$. It seems that the phase transitions are always Ising transitions for any value of $\alpha>0$. 
To confirm this intuitive argument, we will determine the other critical exponent called critical adiabatic dimension $\mu$ in the following section.

{\it Critical exponents.-} It is well known that for Ising model the critical adiabatic dimension $\mu=2$ 
and the critical adiabatic dimension $\nu=1$ \cite{Zanardi2006,Gu2010,Damski2013,Damski2014}.
And from the theory of fidelity susceptibility, the finite-size scaling behavior of fidelity susceptibility is given in Eq.(\ref{eqFSscaling}) for a second-order 
quantum phase transition near the critical point. The finite-size scaling behavior of fidelity susceptibility is shown in Fig.\ref{FSdata} for 
$\alpha=0.8$ and $L=96,144,192,240$ sites. The maximal of fidelity susceptibility per site $\chi_{L}(h)=\chi_{F}(h)/L$ defined in Eq.(\ref{eqFSsite}) 
increases linearly, indicating the critical adiabatic dimension $\mu=2$. As we expected, we found that all the critical adiabatic dimension $\mu=2$
for the whole range of $0<\alpha \le 3$, whose data is plotted in Fig.\ref{Mudata}. Therefore the numerical results from the fidelity susceptibility 
show that the nature of phase transitions are second-order Ising transitions for all the values of $\alpha>0$.

We note that our results are not in conflict with the previous numerical data obtained determined from entanglement entropy \cite{Koffel2012,Vodola2016,Jaschke2017}.
Indeed, the critical values $h_c$ shown here are consistent with previous results obtained from maximal of entanglement entropy \cite{Koffel2012,Vodola2016,Jaschke2017}. 
There may be two possible reasons for the different conclusions for small $\alpha$ between the fidelity susceptibility and entanglement entropy.
One is the numerical reason: The gap is too small and the ground-states are highly entangled for small $\alpha$ so that one needs
large system sizes and large bond dimensions of the matrix blocks to make a correct extrapolation. And for the extrapolation of entanglement entropy,
it is known that it is very difficult to perform a logarithmic scaling for small system sizes.
The other is the physical reason: For small $\alpha$, the long-range interactions may raise the effective dimensionality and lead to the breaking 
of the conformal symmetry \cite{Vodola2016}. The entropy and the correlation will behave in a new manner \cite{Koffel2012,Vodola2016}, which is now an open question.
However we argue that the phase transition can still be the second-order Ising transition corresponding to higher effective dimension for small $\alpha$.

The advantage of the usage of the fidelity susceptibility is that we can use the well established theory of scaling behavior of fidelity susceptibility 
to extract the critical exponents and determine the nature of phase transitions. We agree that
more numerical and analytical methods are in demand in future investigations to confirm previous results and explained the novel behaviors.
Here we provide the very nice data from the concept of fidelity susceptibility for the first time to understand the nature of the LRAI chain.

%%%%%%%%%%%%%%%%%%%%%%%%%%%%%%%
% Conclusion
{\it Conclusion.-} We have investigated the finite-size scaling of the fidelity susceptibility $\chi_{F}$ and the nature of quantum phase transitions of the LRAI chain 
using the DMRG simulations. In particular, we have shown that all the quantum phase transitions are Ising transitions with an Ising universality class for any value of $\alpha>0$.
This study provides a general approach to understand the quantum phase transitions in long-range interacting quantum many-body systems. 
In the future, it would be very interesting to use such techniques to study the nature of quantum phase transitions or to find new exotic quantum phases 
in other quantum many-body systems in the presence of long-range interactions, such as one dimensional long-range interacting Bose-Hubbard model \cite{Jaksch1998}
or two dimensional spin models. Furthermore, it is important to explore an effective theoretical method to explain the nature of phase transitions 
and support our numerical simulations in the highly entangled long-range ($\alpha <1$) regime in future investigations.

% Acknowledgments
\begin{acknowledgments}
G.S. would like to thank T. Vekua for useful discussions and comments for the paper. 
G.S. is appreciative of support from the Start-up Fund of Nanjing University of Aeronautics and Astronautics under the Grant No. 90YAH17053. 
Numerical simulations were done on the cluster of the Max Planck Institute for the Physics of Complex Systems, Germany.
\end{acknowledgments}

\end{document}